\documentclass{article}

\usepackage{arxiv}

\usepackage[utf8]{inputenc} 
\usepackage[T1]{fontenc}    
\usepackage{hyperref}       
\usepackage{url}            
\usepackage{booktabs}       
\usepackage{amsfonts}       
\usepackage{nicefrac}       
\usepackage{microtype}      
\usepackage{cleveref}       
\usepackage{lipsum}         
\usepackage{graphicx}
\usepackage[numbers]{natbib}
\usepackage{doi}
\usepackage{color}

\usepackage{float}

\title{What defines a group of friends? Rethinking community structure in signed, directed networks}

\newif\ifuniqueAffiliation

\ifuniqueAffiliation 
\author{ \href{https://orcid.org/0000-0000-0000-0000}{\includegraphics[scale=0.06]{orcid.pdf}\hspace{1mm}Miguel A. González-Casado}\thanks{Use footnote for providing further
		information about author (webpage, alternative
		address)---\emph{not} for acknowledging funding agencies.} \\
	Department of Computer Science\\
	Cranberry-Lemon University\\
	Pittsburgh, PA 15213 \\
	\texttt{hippo@cs.cranberry-lemon.edu} \\
	\And
	\href{https://orcid.org/0000-0000-0000-0000}{\includegraphics[scale=0.06]{orcid.pdf}\hspace{1mm}Elias D.~Striatum} \\
	Department of Electrical Engineering\\
	Mount-Sheikh University\\
	Santa Narimana, Levand \\
	\texttt{stariate@ee.mount-sheikh.edu} \\
}
\else
\usepackage{authblk}

\author[a,1]{Miguel A. González-Casado}
\author[a,b]{Angel Sánchez}
\author[c]{Santo Fortunato}

\affil[a]{Grupo Interdisciplinar de Sistemas Complejos (GISC), Departamento de Matemáticas, Universidad Carlos III de Madrid, 28911 Legan\'es, Spain}
\affil[b]{Instituto de Biocomputaci\'on y F\'isica de Sistemas Complejos (BIFI), Universidad de Zaragoza, 50018 Zaragoza, Spain}
\affil[c]{Center for Complex Networks and Systems Research, Luddy School of Informatics, Computing, and Engineering, Indiana University Bloomington, USA}
\affil[1]{To whom correspondence should be addressed. E-mail: miguelangel.gonzalezc@outlook.es}

\hypersetup{
pdftitle={TBD},
pdfsubject={TBD},
pdfauthor={TBD},
pdfkeywords={TBD},
}

\begin{document}
\maketitle

\begin{abstract}
We study the structure of personal relationships among 1068 high school students using a dataset that contains the network of self-reported friendly and conflictive relationships, with information on their directionality and intensity. We analyse the resulting weighted, directed, and signed network using a Bayesian stochastic block model framework, which enables the inference of group structure without imposing prior assumptions on the role of negative or asymmetric ties. While a full model incorporating all edge attributes yields statistically coherent clusters, these do not align with socially meaningful communities. To address this, we focus first on the network backbone of mutual affinities, and we characterize its group organization. Many communities display an assortative structure, often embedded within larger cohesive configurations, but we also observe more diverse patterns such as core–periphery structure and isolated nodes. We then examine how relationship intensity, directionality, and conflict shape group structure. Asymmetric ties, though often occurring between communities, are frequently present within them, revealing the stabilizing effect of group membership on non-mutual relationships. Furthermore, the presence of asymmetric ties does not inherently imply a hierarchical structure, given that all groups both receive and report significant levels of non-reciprocal ties. More intense ties play a disproportionate role in shaping community structure. Finally, negative ties tend to bridge communities, but we find that groups feature a significant level of internal conflict. Our research offers a new perspective on the study of group organization when rich information about the directionality, the intensity and the sign of ties is considered, with implications for identifying social vulnerability and designing targeted interventions.
\end{abstract}


\section*{Introduction}
At first glance, asking what defines a group of friends seems like a simple question: we all believe we recognize our own groups of friends, and we share intuitions about who belongs to them and who does not. However, upon closer inspection, the concept becomes elusive. Intuitively, we could define a group of friends as the set of people you would invite to a plan altogether, with whom you would share a dinner, a trip, or a table at a wedding. A group of friends may often arise from shared contexts (classmates, coworkers, or shared activities members), but would rarely include everyone in that context. Its structure and boundaries may not be always clear: they can arise from random interactions that crystallize over time, vary greatly in size, fragment due to conflicts or practical limitations, or evolve in different ways. There is often an assumed affinity among group members, though not necessarily between all of them or to the same degree. Moreover, such affinity may not be always the origin of the group; it may instead emerge as a consequence of coexistence and shared experiences. In any case, the group is an abstract entity: it exists as a shared psychological representation. Previous research shows that, within social groups, individuals tend to be more committed, care about the group, and develop distinctive cultural features such as inside jokes, rituals, and shared practices \cite{ridgeway1983dynamics}. Norms also emerge to regulate members' behaviour, define acceptable interactions, and ensure the group's cohesion as a social unit. Moreover, group members are aware of the group's existence and develop shared perceptions, emotional bonds, and organizational structures in terms of roles, statuses, and norms \cite{shaw1961group}. Because of this inherently abstract nature, empirically studying social groups is a complex task. 

From a network-theoretical perspective, a common intuition is that members of the same group share more connections with each other than with members of other groups and therefore social groups are network communities~\cite{wasserman94,newman2004finding, porter09,fortunato2010community,fortunato16,fortunato22}. Consequently, uncovering group structure from a network of personal relationships would amount to detecting communities. However, this does not take into account the complexity of real-world patterns. In practice, group structures can exhibit a wide spectrum of behaviours that go far beyond the traditional concept of assortative community above. Some individuals may act as satellites of cohesive groups, maintaining ties only with a subset of members; small cores or individuals may simultaneously belong to several larger groups; and transient configurations may arise in which groups aggregate, fragment, or merge. This complexity is further amplified by the fact that personal relationships differ in intensity, may be asymmetric, conflictive, and evolve continuously over time. Therefore, one could, in principle, define a `group' in many different ways depending on the relational structure and the available information about tie strength, directionality, and conflict. Each definition implies specific assumptions about how network structure reflects underlying group organization.

Incorporating information about the intensity, directionality, and conflict of relationships adds layers of complexity that can be difficult to handle when identifying groups. For instance, previous studies have shown that friendships of different strengths can be driven by distinct underlying mechanisms~\cite{leenders1996evolution}. Consequently, integrating this information into a community detection framework is not straightforward, specially when the research question concerns the nature of the communities identified and the role played by relationship intensity. Moreover, many real-life relationships are asymmetric, and strong domain-specific assumptions about how such ties emerge may help to find the communities~\cite{feuer2022reciprocity}. Yet, empirical evidence on where asymmetric ties appear and what role they play in shaping community organization remains limited. The situation becomes even more complex when conflicting ties are considered. Most community detection algorithms, influenced by balance theory \cite{cartwright1956structural,doreian2009partitioning}, assume that negative edges primarily occur between groups, while positive edges dominate within groups. This assumption can be operationalized through frustration minimization~\cite{bansal04} or by modifying the modularity function~\cite{gomez09,traag2009community}. However, the goal of the present work is precisely to analyze these behaviors without imposing such a priori assumptions, as we shall discuss.

Only a few studies have analyzed the group structure of friendship networks while incorporating information on relationship intensity, directionality, and conflictive ties. They adopt very different methodological frameworks and assumptions, making direct comparison difficult. Examples include \cite{escribano2021evolution}, \cite{brito2022modularity} and \cite{stadtfeld2020emergence}, which examined the group structure of high school social networks while accounting for negative ties. However, their approaches differ substantially from ours, as they explicitly incorporate strong assumptions about the behaviour of negative edges consistent with predictions from balance theory, as discussed earlier. It is worth mentioning that, within the context of friendship networks, some authors have adopted a complementary, more mechanistic approach to the study of group structure. Rather than simply describing the observed network, this approach seeks to explain the emergence of collective patterns from basic microscopic social behaviours. For example, individuals, constrained by limited time and resources, tend to bring their friends together~\cite{feld1981focused}. Early works tried to formalize this type of intuition through computational models~\cite{zeggelink1996emergence}. \cite{block2018network} explored how three fundamental mechanisms (reciprocity, transitivity, and homophily) interact to produce complex structures. They found that the competition between homophily and transitivity can generate overlapping group memberships, making individuals bridge between different network regions and giving rise to non-trivial group configurations beyond classical modular structures. Beyond these endogenous mechanisms, network structure is also shaped by contextual and institutional factors that can act as exogenous mechanisms. For instance, \cite{moody2001race} demonstrated that racial segregation in U.S. schools largely results from organizational constraints that govern student interactions. Within the framework of network ecology \cite{doehne2024network}, such constraints reflect the social histories and normative `network institutions' that define acceptable relational behaviour, channeling individual choices into a limited set of viable connections. Consequently, different contexts develop distinct relational norms that shape group structure. \cite{mcfarland2014network} further showed that while microscopic mechanisms such as reciprocity or transitivity operate universally, their combination with these ecological constraints, such as demographic composition, institutional structure, or educational climate, produces context-specific mesoscopic organizations. In other words, the same underlying principles can lead to different group structures depending on the social environment in which they unfold.  

All in all, the question we ask ourselves is: what insights about group structure can we get from the relational structure of a friendship network that includes information about weights, directions and signs? Drawing on more than five years of data on the structure and dynamics of networks of personal relationship between groups of adolescents, we begin with a detailed analysis of the structure of the community. We characterize the different types of community organization displayed in the data, disentangling modular structures from core–periphery and other complex patterns, while also examining the role of group size and other general characteristics of groups. In particular, we show that some community detection methods applied blindly can mix different types of assortative patterns, specially when incorporating information about the direction, weight and sign of the edges, and thus we show the need for careful, detailed interpretation to avoid misleading conclusions. 
We then take advantage of the richness of our dataset, which captures not only the intensity of relationships but also their directionality (as relationships are self-reported) and the presence of negative or conflictive ties. The resulting network is directed, weighted, and signed, with weights ranging from –2 to +2. Communities are detected via statistical inference of stochastic blockmodels (SBMs)~\cite{peixoto20}. To our knowledge, our work features the first thorough analysis of communities in dynamic social networks with directed, weighted, and signed edges, all features that can be taken into account by suitable SBMs, whereas most community detection approaches are not capable of dealing with them.

After the groups in different snapshots of the evolving networks are identified, we examine the role of ties of different intensities in shaping group structure, exploring where asymmetric ties occur, whether they signal the presence of hierarchical relations, and whether they appear predominantly within or between groups, and in what ways. Furthermore, given the signed nature of our data, we investigate the placement and role of negative relationships in group-level structures, testing long-standing hypotheses from structural balance theory, including the idea that conflicts tend to occur between, rather than within, groups.

\section*{Results}

In this work, we analyse a dataset we have collected between 2020 and 2025, that contains the temporal evolution of the network of personal relationships among 1068 students belonging to a High School in Madrid (see the Methods section for details on the data collection, composition, and curation). The reported relationships were coded as +2 (very good), +1 (good), -1 (bad), and -2 (very bad). Almost every student provided this list, resulting in the extraction of a weighted directed network for the entire high school, which we will refer to as snapshot or wave. We repeated this process every 16-20 weeks, and with this data we reconstructed the network at different points in time, collecting 14 snapshots of the network over 5 years. Preliminary analyses of earlier, shorter versions of this dataset have been reported elsewhere \cite{escribano2023stability,gonzalez2025evidence}. 

For each snapshot we have a directed, weighted, and signed network. The purpose of our analysis is to study the group organization underlying the observed structure of personal relationships, and for that we apply a community detection algorithm. The first problem is choosing a suitable algorithm. As we specified in the introduction, our purpose is exploring community organization beyond simple assortative structures. Modularity maximization, by far the most common approach, besides its well-known shortcomings \cite{fortunato07,peixoto2023descriptive}, is unsuitable for the goals of this paper for two other reasons. First, defining groups solely as assortative structures is an oversimplification compared to the range of group patterns we expect to find in real-world networks. We aim for a community detection method that can reveal not only assortative structures, but also core–periphery structures and other types of mixing patterns. Second, our data include edges with varying weights, signs, and directions. While edge weights and directions are usually not problematic, signed edges are tricky to handle. 
Balance theory suggests that signed edges should preferentially be found between groups, but we will not make that assumption here, as 
it is entirely possible for numerous conflicts to exist within a group of friends whose positive relationships still define them as a cohesive unit. 

We employ a nonparametric Bayesian inference approach based on the family of Stochastic Block Models (SBMs) (see Methods section). This method naturally accommodates a broad range of structures beyond standard assortative communities, including disassortative communities, core–periphery structure, and combinations thereof. The strategy we follow is straightforward. First, we find the best model for the data, which is the one with the minimum description length. This provides us with the partition of nodes that best explains the observed data, given the assumptions of the model. We then use this partition together with the network’s adjacency matrix to analyse in detail the structural patterns of the partition and to understand: (1) what the algorithm identified as groups, (2) why it identified them based on the observed network structure and (3) what structural insights can be drawn regarding the behaviour of other network features.

Given that we have detailed information on the directions, weights, and signs of the network ties, and that our framework is flexible enough to incorporate all of this information to infer communities, the most natural strategy is to include it in full. However, as we demonstrate in Section A of the Appendix, this strategy is not suitable for our research goals. If our objective is to use a rigorous approach for the identification of communities in a networked structure and give them a social interpretation as outlined in the introduction (groups shaped by personal affinities and shared membership as they are intuitively understood), then we cannot simply fit a stochastic block model using all available network features and assume that the resulting clusters correspond to meaningful social groups. The need for caution becomes evident once we examine the structural diversity that the algorithm actually detects. As illustrated in Section A of the Appendix, the SBM identifies a wide variety of connectivity patterns. While some inferred communities do match an intuitive notion of a social group (for instance, clusters characterized by dense, reciprocal positive ties) others reflect very different structural logics. Some communities are defined primarily by internal antagonism, grouping together nodes that share dense patterns of negative ties. Others are organized around conflictive hubs, nodes that report large numbers of negative ties both inside and outside the community, meaning that what binds these nodes together is not mutual interaction but statistically similar patterns of outward hostility. Others correspond to sociable or popular hubs: nodes that report or receive many positive ties but have sparse internal connections, making them structurally cohesive only in the statistical sense captured by the algorithm, not in the social sense that motivates our research questions. These examples show that the algorithm is functioning exactly as designed: it clusters nodes according to statistically similar connectivity patterns. The challenge is not methodological but interpretive. Community detection algorithms, especially when applied to weighted, signed, and directed networks, reveal a spectrum of structural behaviours, some intuitive, others highly non-intuitive from a social-group perspective. Yet our research questions, intuitions, and interpretations often default to reading any detected `community' as a social group. The more dimensions of network information we include (direction, weights, signs), the greater this mismatch becomes, because the resulting structural groupings increasingly reflect heterogeneous motifs that do not align with the kinds of socially cohesive groups we aim to study. For this reason, relying on an unconstrained SBM fitted directly to the full set of network features would obscure rather than clarify the questions of interest. This result motivates the reframing of our methodological strategy to one that makes the detection and interpretation of groups more transparent and more closely aligned with our substantive goals.

\subsection*{Groups defined by reciprocal relationships}
\begin{figure}[t!]
\centering
\includegraphics[width=\linewidth]{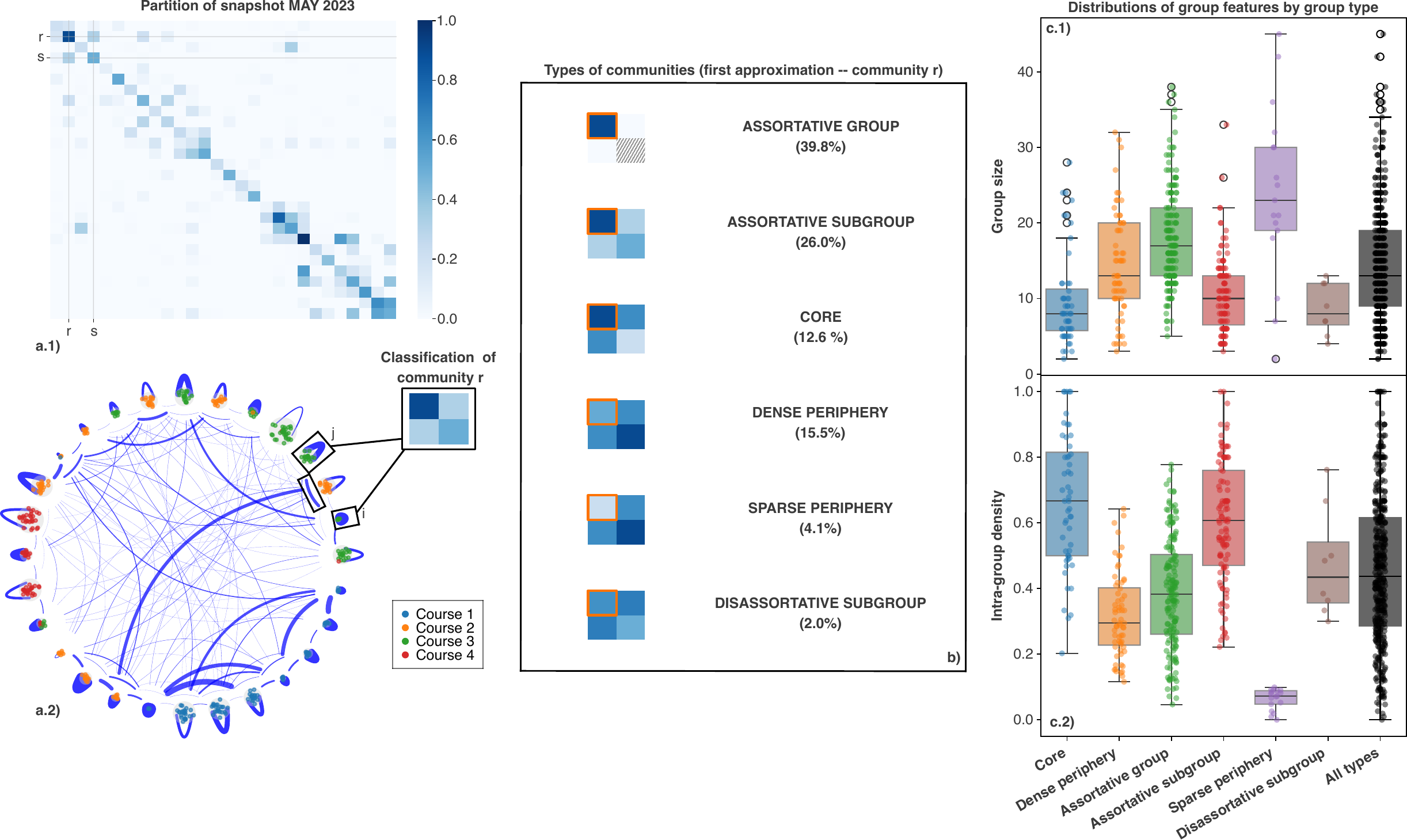}
\caption{\footnotesize \textbf{Characterization of the simplified networks' community structure}. \textbf{Panel a.1)} Edge count matrix $e$ for the May 2023 snapshot, used here as an example. Each row and column corresponds to a community detected by the algorithm, and each element $e_{rs}$ represents the density of edges between groups $r$ and $s$ (self-loops included), normalized by the maximum number of potential edges between those two groups. \textbf{Panel a.2)} Network representation of the matrix in panel a.1). Here nodes represent communities, and edge width is proportional to the weight in element $e_{rs}$ of the matrix. Inside each node, we display as many coloured points as individuals belong to that community, to illustrate differences in community size. The colour of the points indicates the school course of the individuals. Two communities, $r$ and $s$, are highlighted to exemplify the classification procedure summarized in panel b) and described in the main text. \textbf{Panel b)} Schematic representation of the different approximate behaviours community $r$ may display depending on the values of $e_{rr}$, $e_{r\bar{s}}$, and $e_{\bar{s}\bar{s}}$, where $\bar{s}$ is the community with the strongest connection to $r$. For each type of community, we include the percentage of communities in our dataset displaying that behaviour. \textbf{Panel c.1)} Distribution of community sizes, both aggregated and separated by group type. \textbf{Panel c.2)} Distribution of intra-group densities, both aggregated and separated by group type.} 
\label{fig2}
\end{figure}

It is considered essential for social relationships that sentiments are reciprocated \cite{kitts2021n,ready2021measuring}, and it has been empirically found that they often are \cite{rivera2010dynamics}. Furthermore, it has been shown that asymmetric relationships tend to either dissolve or develop into mutual ties \cite{hallinan1978process}. At the group level, \cite{coleman1961theadolescent} recognises the importance of reciprocal relationships in the formation of friendship groups, defining a group member as someone who has reciprocal relationships with at least two others in the group, and \cite{feuer2022reciprocity} stresses that, in friendship networks, one important network structure that aids at identifying communities is reciprocal relationships. Building on these works, based on the intuition that a group, in the social sense, is defined by personal relationships and mutual affinities among its members, in our new approach we extract a projected backbone of the network. Specifically, we simplify the network by retaining only reciprocal positive relationships, yielding a simple, undirected, unweighted, and unsigned graph. Community detection is then applied to this simplified representation, where the rationale for why the algorithm identifies particular communities becomes more straightforward and transparent. Once this baseline community structure is characterized, we subsequently reintroduce the information on directions, weights, and signs, and analyse how these features behave on top of the pre-specified group structure. In this way, we can study how asymmetric or conflictive ties map onto a group structure that is already aligned with our social intuitions. 

The projection is constructed by removing all negative edges, as well as positive edges that are not reciprocated (i.e., asymmetric ties). This leaves three types of reciprocal positive edges: (i) mutual ties of weight +2, (ii) mutual ties of weight +1, and (iii) mixed ties where one node reports +1 and the other reports +2. In the subsequent analyses, we examined a projection in which all three types of reciprocal positive edges were retained and assigned the same weight. Retaining only the +2 reciprocal edges, by contrast, results in a network that is overly sparse and exhibits a fairly trivial community organization. Importantly, while the three types of reciprocal edges are treated as equivalent during community detection, we subsequently reintroduce their weights, alongside the asymmetric and negative ties, when analysing the structure. This allows us to study not only the placement of asymmetric and conflictive ties, but also the role of relationship intensity within the previously identified group structure. 

In our projection we obtain a simple, unweighted, undirected, and unsigned network, which now we need to fit using some variant of the SBM. For that we need to identify which features should be incorporated into the different SBM variants considered. Of course, we include the two features introduced in the preliminary analysis reported in Section A of the Appendix (nestedness and degree correction) as they are expected to remain relevant in the projected networks as well as in the original one. In addition, it is informative to consider the planted partition model, a special case of the SBM that imposes communities to be strictly assortative or disassortative (see Methods). This version restricts the inferred structures to purely assortative partitions, which may in fact be appropriate for describing group structure in this simplified representation of the network. Finally, as also discussed in the Methods section, we must account for the possibility that triadic closure (the mechanism by which edges are more likely to form between individuals who share a common neighbour) plays a role. Triadic closure can give rise to apparent block structures even when they do not reflect meaningful communities in an inferential sense. In such cases, a standard SBM might detect block partitions that are in fact artifacts of this local mechanism. To evaluate this possibility, we incorporate a model that explicitly accounts for triadic closure and examine whether it provides a better description of the observed structure. Taken together, we therefore consider the following set of models:
\begin{itemize}
    \item Traditional SBM 
    \item Degree-Corrected SBM
    \item Nested SBM  
    \item Nested Degree-Corrected SBM 
    \item Planted Partition SBM
    \item Traditional SBM with Triadic Closure
    \item Degree-Corrected SBM with Triadic Closure
    \item Nested SBM with Triadic Closure
    \item Nested Degree-Corrected SBM with Triadic Closure
\end{itemize}
Interestingly, in both projections and across all 14 network snapshots, the model that consistently yields the partition with the minimum description length is the Nested SBM variant. Notice this model is nested, as expected, but not degree-corrected. This result also indicates that the detected group structures cannot be reduced to purely assortative communities; instead, they must involve more complex patterns, which we explore in detail in the following section. Furthermore, although triadic closure is likely to play a role in shaping the network, incorporating it into the model does not improve the description of the observed group structures. With these results in mind, we now proceed to analyse the structure of the partitions that minimized description length in the optimal model for both projections.

At this stage of the analysis we are not concerned with the dynamical evolution of the network. Our current objective is to characterize in detail the types of group patterns detected by the algorithm; the analysis is therefore purely structural. For this reason, we aggregate the results from all 14 waves and analyse them jointly. While it is possible that some behaviours change as the network evolves, this is not the focus here. Moreover, as shown in \cite{gonzalez2025evidence}, this network displays dynamics consistent with equilibrium: all snapshots are statistically equivalent, and thus we do not expect that the behaviour of mesoscopic properties evolves systematically over time.

For each network snapshot we obtain a partition into communities (here we always refer to the undirected, unweighted, and unsigned version of the network). Using the network's adjacency matrix, we can construct the edge count matrix, which serves as the basis for our characterization of the partitions. If the partition consists of $B$ groups, the matrix of edge counts $e$ is a $B \times B$ matrix where each entry $e_{rs}$ represents the number of edges between groups $r$ and $s$. To properly account for group size, we use a normalized version of this matrix, in which each entry is divided by the maximum possible number of edges for that pair. In this way, the normalized $e_{rs}$ can be interpreted as a density, producing a sort of `supra-adjacency matrix' where each node is a group and self-loops are allowed.

As noted in the Methods, the SBM imposes no restrictions on the type of block structure, so the edge count matrix is unconstrained. It can therefore encode a variety of patterns (assortative, disassortative, core–periphery, and combinations thereof). Given this matrix, we want to determine the nature of each group and understand why the algorithm identified it as a community. To fully characterize a group $r$, one would need to examine $e_{rr}$ (edges within $r$) together with all $e_{rs}$ (edges from $r$ to every other group). For example, if $e_{rr}$ is large while all $e_{rs}$ are small, $r$ is a typical assortative community. But if $e_{rs}$ is large for some $s$, $r$ may in fact constitute the periphery of group $s$. The scenario is more complex when $r$ shows both a strong internal edge density and significant connections to multiple other groups also strongly connected among themselves.

To simplify the problem, we adopt an approximation. For each group $r$, we focus only on three entries of the edge count matrix: $e_{rr}$, $e_{r\bar{s}}$, and $e_{\bar{s}\bar{s}}$, where $\bar{s}$ is the group to which $r$ is most strongly connected (i.e., the $s$ that maximizes $e_{rs}$). The rationale is that, as a first approximation, the structural role of $r$ can be characterized by looking at its internal density and the group it is most tightly connected to. In general, if communities tend to be highly interconnected and community $r$ maintains strong ties with several others, this would be a poor approximation, since focusing only on a subset of its connections would ignore much of the relevant structure. However, if community $r$ concentrates most of its edges internally and with just one or a few other communities, then characterizing it using only $e_{rr}$, $e_{r\bar{s}}$, and $e_{\bar{s}\bar{s}}$ offers a reasonable approximation. In Section B of the Appendix, we show that the majority of communities exhibit this pattern, maintaining most of their ties internally and with the group they are most tightly connected to, instead of distributing them evenly among all external connections, thus supporting the validity of this simplification.

The consideration of $e_{rr}$, $e_{r\bar{s}}$, and $e_{\bar{s}\bar{s}}$ yields a useful set of diagnostic patterns. Just by considering these three values, we can classify the behaviour of community $r$ into three types. If $e_{r\bar{s}}$ is smaller than both $e_{rr}$ and $e_{\bar{s}\bar{s}}$, we identify the typical assortative structure, where $r$ is mostly connected to itself. If instead $e_{r\bar{s}}$ lies between $e_{rr}$ and $e_{\bar{s}\bar{s}}$, we observe a core–periphery structure: if $e_{rr} > e_{\bar{s}\bar{s}}$, $r$ is the core; if $e_{rr} < e_{\bar{s}\bar{s}}$, $r$ is the periphery of $\bar{s}$. Finally, if $e_{r\bar{s}}$ is larger than both $e_{rr}$ and $e_{\bar{s}\bar{s}}$, we are in a situation where inter-group connections dominate over intra-group ones, resembling a disassortative structure in which both groups are more embedded in one another than in themselves. This criterion suffices to classify community $r$ within our approximation. However, the purpose of our classification is not merely to describe structural patterns, but to extract socially meaningful insights from them. To do so, we introduce further subdivisions. In the case where $e_{r\bar{s}}$ is smaller than both $e_{rr}$ and $e_{\bar{s}\bar{s}}$, corresponding to a pair of assortative communities, it is socially relevant to distinguish between the case in which $e_{r\bar{s}}$ is very small, indicating two isolated assortative communities, and the case in which $e_{r\bar{s}}$ is substantially larger (though still smaller than $e_{rr}$ and $e_{\bar{s}\bar{s}}$), suggesting that both groups are assortative subgroups embedded within a larger cohesive structure. This distinction, although arbitrary, becomes meaningful in contexts such as conflict analysis: a conflict between two isolated groups is not the same as one between subgroups within a broader structure of positive ties. To differentiate these cases, we impose an arbitrary threshold of $0.2$ on $e_{r\bar{s}}$. Changing this threshold does not affect our conclusions, as it serves only as a tool for interpretation. Similarly, when $e_{rr} < e_{r\bar{s}} < e_{\bar{s}\bar{s}}$, such that $r$ is classified as the periphery of $\bar{s}$, it is again relevant to distinguish between a periphery that is densely connected internally (where $e_{rr}$ is still relatively large), and one where $e_{rr}$ is close to zero. In the latter case, $r$ may not constitute a community in the social sense, but rather a set of nodes grouped together due to similar external connectivity patterns (e.g., nodes that are structurally isolated). In contrast, when $e_{rr}$ is large, we may have identified a cohesive structure consisting of a central cluster (core) and a more loosely connected periphery that nonetheless belongs to the same group. To differentiate between dense and sparse peripheries, we apply the same threshold on $e_{rr}$. With these two subdivisions, we arrive at a classification of community $r$ into six categories: assortative groups, assortative subgroups, cores, dense peripheries, sparse peripheries, and disassortative subgroups.

Within this framework, we can now fully characterize the communities identified. In Figure~\ref{fig2} we show the proportions of all characteristic group patterns, along with the distributions of sizes and densities for each type of group. We find that the majority of communities are assortative groups (40\%), largely isolated from the rest of the network. A further 26\% of communities correspond to assortative subgroups embedded within a larger cohesive structure. Altogether, about two thirds of the groups found correspond to intuitive assortative communities, either isolated or embedded to some extent within larger, sparser structures. In addition, 19.6\% of the groups are identified as peripheries of another group. Most of these (15.5\%) are densely connected, while the remaining 4.1\% are sparsely connected. These sparse peripheries are better interpreted as sets of nodes that are relatively isolated but attached to some other (denser) structure, which the algorithm groups together. Finally, 2\% of the groups correspond to cases where all $e_{r\bar{s}}$, $e_{rr}$ and $e_{\bar{s}\bar{s}}$ are large, with $e_{r\bar{s}}$ only slightly larger than the other two. Strictly speaking, these resemble a structure closer to dissasortative, in the sense that more connections occur between the two groups than within them, even though connections are significant both within and between the two groups. Turning to group size and density, the median size across all groups is about 12 individuals, with most groups ranging between 9 and 20 members, although the full distribution spans from 2 to 40. Cores and assortative subgroups are small and very dense, while the disassortative subgroups are slightly sparser but still fit within this pattern. Cohesive subgroups tend to be somewhat larger and consequently less dense, and the same applies to dense peripheries. In contrast, sparse peripheries are the largest and, as their name suggests, the sparsest. 

Although approximate, this analysis provides a detailed characterization of the group structures in these social networks. Most groups are cohesive communities, to varying degrees embedded within larger, sparser structures. This embeddedness naturally gives rise to peripheral structures when some nodes fail to aggregate strongly among themselves.

\subsection*{Non-reciprocal edges: in-group vs out-group asymmetries and hierarchies}
\begin{figure}[t!]
\centering
\includegraphics[width=\linewidth]{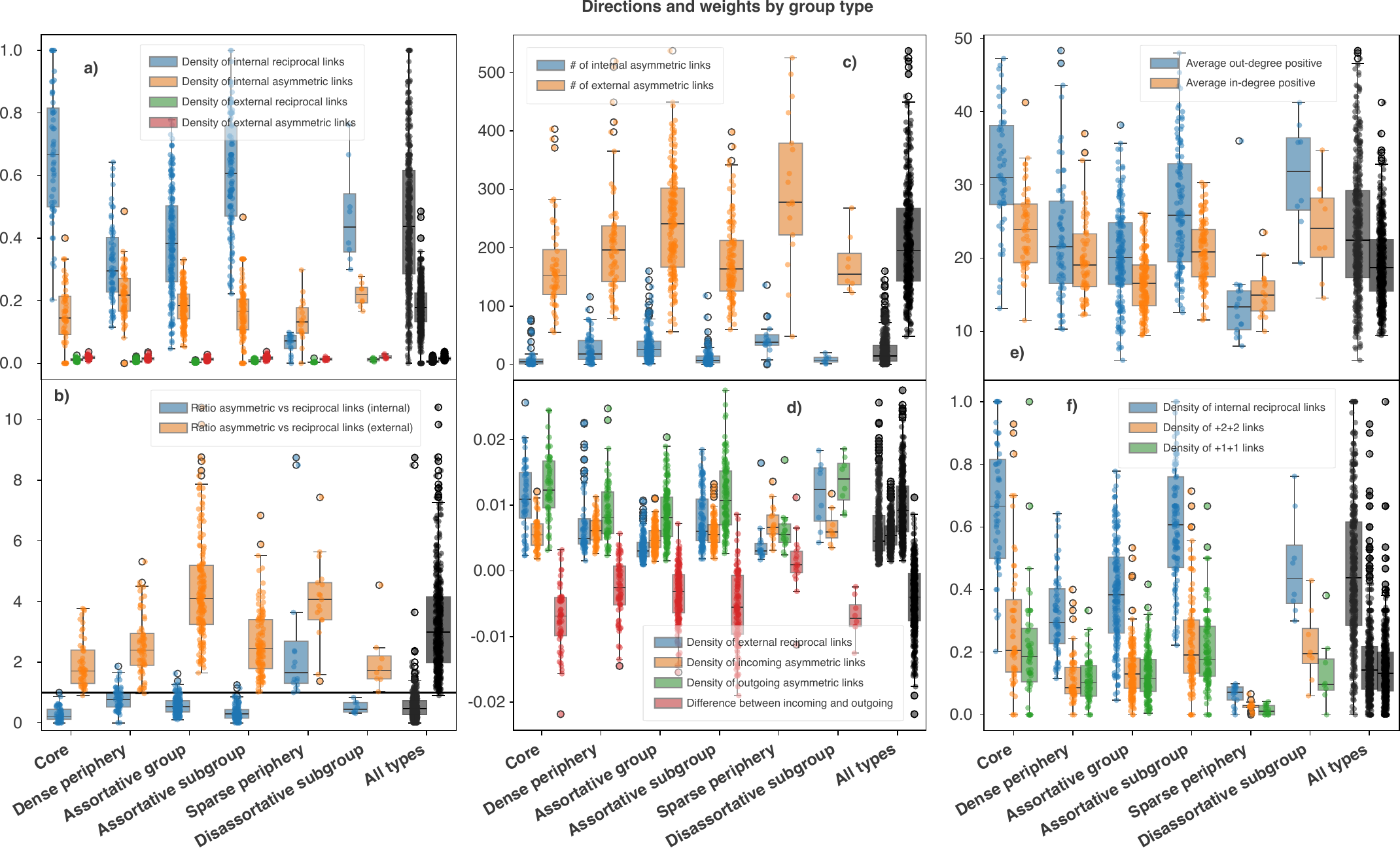}
\caption{\footnotesize \textbf{Characterization of the behaviour of asymmetric positive edges and the role of positive edge intensity.} \textbf{Panel a)} Distributions of four densities: intra-group and inter-group densities of both positive reciprocal and positive asymmetric edges for each community. \textbf{Panel b)} Distributions of the ratio between the density of positive asymmetric edges and the density of positive reciprocal edges, for both intra-group and inter-group ties. \textbf{Panel c)} Raw counts of intra-group and inter-group positive asymmetric edges. \textbf{Panel d)} Distributions of three densities: inter-group density of positive asymmetric edges (combining incoming and outgoing), and inter-group densities of incoming and outgoing asymmetric edges considered separately. Additionally, the distribution of the difference between incoming and outgoing densities for each community is shown. \textbf{Panel e)} Distributions of the average positive in-degree and out-degree for each community. \textbf{Panel f)} Distributions of three densities: intra-group density of positive reciprocal edges (irrespective of weight), and intra-group densities of reciprocal $+2+2$ and $+1+1$ edges separately. All panels display the distributions, both aggregated and separated by group type.} 
\label{fig3}
\end{figure}
So far, we have characterized the group structure in the projection of the network composed of reciprocal positive relationships (both +1 and +2 relationships, treated equivalently), assuming that the structure of reciprocal affinities serves as the backbone of the network's community organization. Now, taking advantage of the richness of our data, we explore more complex behaviours on top of this fundamental group structure. The strategy to do so is to maintain the previously defined group partition, and reintroduce progressively information on directions, weights, and signs of the edges, to extract conclusions about their behaviour. We will start by characterizing the behaviour of asymmetric positive edges and, afterwards, explore briefly the role of edge intensity (+2 vs +1 edge weights).  Thus, as a first step we reintroduce asymmetric positive edges while ignoring their specific weights. Our aim is to compare the behaviour of asymmetric and reciprocal positive ties independently of their intensity. While differences may exist between asymmetric ties of weight $+1$ and $+2$, this level of detail is not necessary for the conclusions we draw, as will become clear below. For each community, we measure some metrics: raw counts of intra-group and inter-group positive asymmetric edges; intra-group and inter-group densities of both positive reciprocal and positive asymmetric edges; the ratio between the density of positive asymmetric edges and the density of positive
reciprocal edges, for both intra-group and inter-group ties; inter-group densities of incoming and outgoing asymmetric edges considered separately; the difference between incoming and outgoing densities; and the average positive in-degree and out-degree. Then, considering all communities, we depict in Figure~\ref{fig3} (panels a)-e)) the distributions of these different metrics, separated by group type according to the classification of the previous section. 

First, we aim to address two questions: (1) do asymmetric ties tend to appear more often inside or outside groups, and in what proportion? and (2) how frequent are they, both inside and outside groups, compared with reciprocal ties in the corresponding settings? Notice that, in both cases, the answer depends on the type of group under consideration. Panels a), b), and c) focus on intra-group and inter-group densities of positive reciprocal and positive asymmetric ties, on the ratio of asymmetric to reciprocal ties inside and outside groups, and on the raw counts of internal and external asymmetric ties. When comparing blindly densities of internal versus external asymmetric ties, we observe that the density of external asymmetric ties is significantly smaller. However, this does not imply that most asymmetric ties are placed within groups. Rather, this is a consequence of using densities, which are normalized by the number of potential ties: the pool of potential external ties is far larger. In contrast, the raw counts clearly show that external asymmetric ties vastly outnumber internal ones. This result is perhaps unsurprising, but there are more conclusions to be extracted.

Looking only at the amount of asymmetric ties can be misleading. Certain group types may appear to have fewer external asymmetric ties simply because they have fewer external ties overall. To control for this, panel a) includes densities of both internal and external reciprocal ties, and panel b) shows the ratios of asymmetric to reciprocal ties. Reciprocal ties serve as a baseline against which to interpret the behaviour of asymmetric ties. Extending the above result, the ratio of asymmetric to reciprocal ties confirms that the ties between the groups are predominantly asymmetric, whereas the ties within the groups are predominantly reciprocal. The exceptions are sparse peripheries, where the notion of an `interior' of the group is not meaningful, since they are essentially collections of unconnected nodes grouped together by the algorithm. As expected, the ratio of external asymmetric ties is smaller for cores, as well as for dense peripheries and assortative subgroups. By definition, these groups are embedded within larger cohesive structures and therefore present more external reciprocal ties. Cohesive subgroups, on the other hand, exhibit the highest ratios of external asymmetric ties. Sparse peripheries, although also embedded within larger structures, display large ratios of external asymmetric ties, largely as a consequence of their low out-degree. We explore these results in more detail in the following discussion.

Turning to the ratio of internal asymmetric to reciprocal ties, the behaviour is particularly striking. Across all groups, the median density of internal reciprocal ties is about 0.45, while the median density of internal asymmetric ties is around 0.2. In other words, although groups are typically dominated by reciprocal ties (except sparse peripheries), nearly one third of the ties within groups are asymmetric. Breaking this down by group type, cores and assortative subgroups exhibit the lowest ratios of internal asymmetry (around 0.25), whereas cohesive modular groups have ratios around 0.5, and dense peripheries approach ratios of one. This highlights an important finding: groups sustain a substantial number of asymmetric ties, and in some cases asymmetric and reciprocal ties are equally frequent. The common assumption that groups are populated simply by mutual affinities is therefore misleading. In reality, groups support a large share of asymmetric relationships. Even when reciprocal ties remain more frequent and their structure identifies a group as such, asymmetric ties often occur in equal measure. This generalizes to the group level a phenomenon previously observed at the triadic level \cite{block2015reciprocity}: groups can stabilize asymmetric ties. In simple terms, an unreciprocated relationship outside of a group is typically unstable \cite{hallinan1978process} (one person notices the imbalance, and the tie either becomes reciprocal or disappears). Within a group, however, social pressures and repeated interactions create a different dynamic. Individuals may behave politely or `nicely' toward others without considering them close, while the other person may interpret this as the existence of a genuine tie. Because group membership ensures sustained interaction, these asymmetric ties are less likely to dissolve and can persist over time.

Moving forward, the abundance of asymmetric ties both within and between groups naturally raises the question of whether these ties signal the presence of hierarchies, or whether they arise randomly. Here we refer to hierarchies at the group level, meaning that some groups tend to report many more ties than they receive, or vice versa. This is distinct from individual-level hierarchies, which concern asymmetries in the ties of single individuals. Panel d) of Figure~\ref{fig3} shows the distributions of the density of outgoing and incoming asymmetric ties for each group type, as well as the distribution of their difference per group. For reference, the density of external reciprocal ties is also included. The results show that, with the exception of sparse peripheries, most groups tend to report more ties than they receive, so the difference between outgoing and incoming ties is generally negative. Of course, some groups in the upper tail of the distribution have positive values, meaning they receive more than they report. These groups occur across all types and may correspond to communities positioned at the top of a hierarchy, characterized by being reported as recipients of positive ties that they do not reciprocate. Most other groups, however, show the opposite pattern: they report more relationships with others than they receive. Sparse peripheries are different. They tend to receive more ties than they report, and their distribution of differences is largely positive. This is better understood by looking at panel e) of Figure~\ref{fig3}, which depicts average in- and out-degree for positive ties per group. Members of sparse peripheries report significantly fewer ties, having lower average out-degree. Their tendency to receive slightly more than they report therefore does not necessarily signal hierarchy, but rather a consequence of their generally low degree. For the other group types, the difference between outgoing and incoming ties is relatively small. Thus, while a small fraction of groups may occupy positions at the top of a hierarchy defined by incoming–outgoing asymmetries, these are not systematically related to group type and remain a minority. Most groups tend to report more ties than they receive. Sparse peripheries present what could be called a pseudo-hierarchical behaviour, but this is induced by their low activity rather than true centrality.

Finally, we briefly comment an interesting result on the role of relationship intensity. Panel f) of Figure~\ref{fig3} compares the densities of reciprocal +1 and reciprocal +2 ties, with the density of all internal reciprocal ties shown as reference. Interestingly, both distributions are very similar. This is striking because, at the level of the whole network, +1 reciprocal ties are nearly five times more frequent than +2 reciprocal ties \cite{gonzalez2025evidence}. This implies that, even though tie strength was not used to infer the communities, reciprocal +2 ties play a disproportionately important role in shaping community structure, while both +1 and +2 reciprocal ties are necessary to sustain a non-trivial organization of the network.

Taken together, these results provide an intuitive explanation for the role of asymmetric ties. Individuals differ in their personal thresholds for considering a relationship to exist, making the line between `tie' and `no tie' blurry. Inside groups, the environment of mutual niceness sustains many asymmetric ties: one individual may consider a relationship to exist, while the other does not, yet the tie persists due to repeated interaction. Outside groups, the larger number of asymmetric ties likely reflects exploratory attempts to connect beyond the group, interactions that are more transient and less clearly defined. Finally, a small fraction of asymmetric ties may reflect hierarchical positioning of certain groups, but these are the minority. In contrast, the analysis of relationship intensity shows that both weak and strong reciprocal ties shape community structure, with stronger ties playing a relatively greater role compared to the network as a whole.

\subsection*{Negative edges and social balance: in-group vs out-group conflicts and hubs}

\begin{figure}[t!]
\centering
\includegraphics[width=\linewidth]{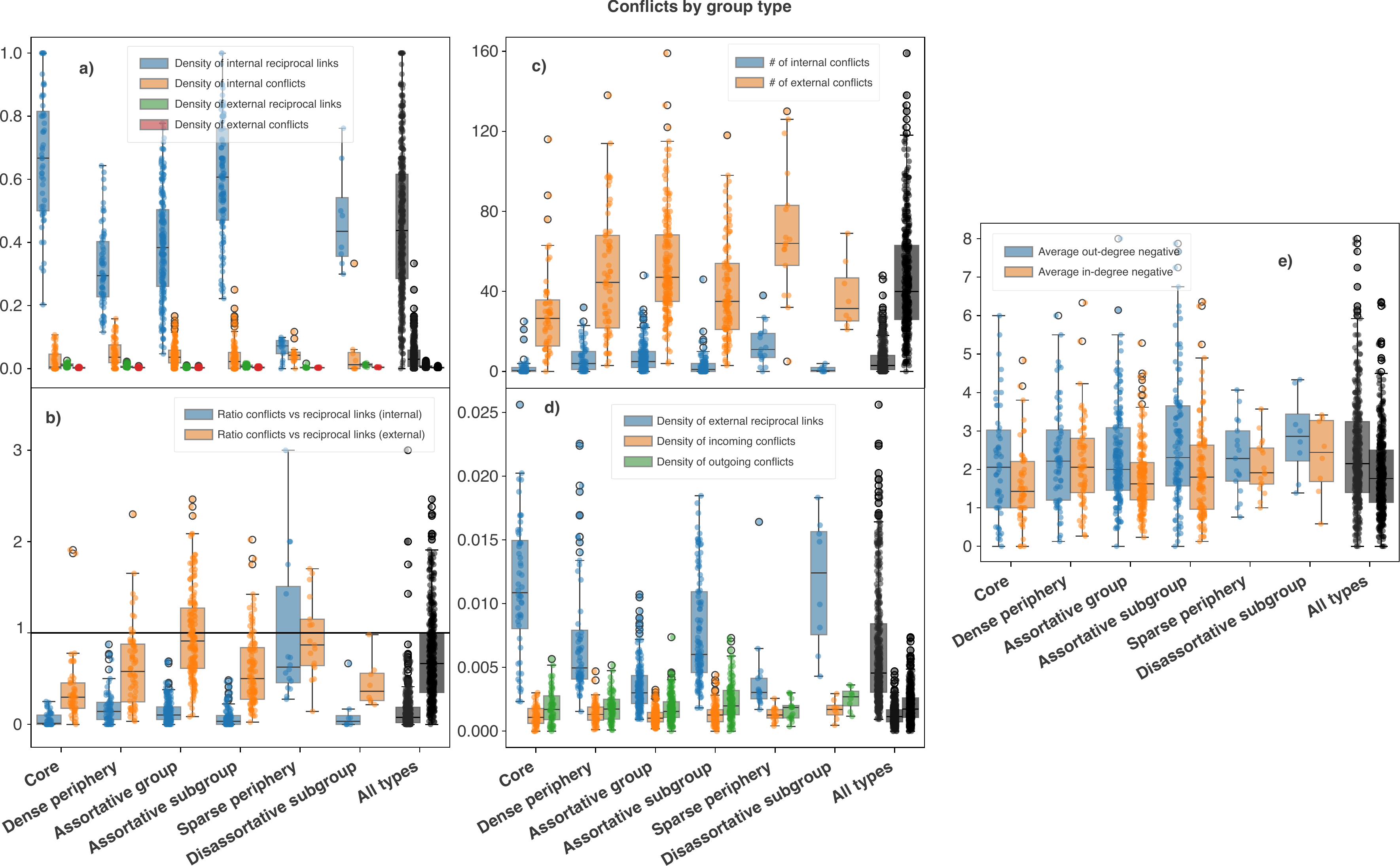}
\caption{\footnotesize \textbf{Characterization of the behaviour of negative edges.} \textbf{Panel a)} Distributions of four densities: intra-group and inter-group densities of both positive reciprocal and negative edges for each community. \textbf{Panel b)} Distributions of the ratio between the density of negative edges and the density of positive reciprocal edges, for both intra-group and inter-group ties. \textbf{Panel c)} Raw counts of intra-group and inter-group negative edges. \textbf{Panel d)} Distributions of three densities: inter-group density of negative edges (combining incoming and outgoing), and inter-group densities of incoming and outgoing negative edges considered separately. \textbf{Panel e)} Distributions of the average negative in-degree and out-degree for each community. All panels display the distributions both aggregated and separated by group type.} 
\label{fig4}
\end{figure}

In the previous section, assuming that reciprocal affinities form the backbone of the network’s community organization, we reintroduced asymmetric positive edges and their weights on top of the pre-specified community structure, and characterized their behaviour. We now follow a similar strategy for negatively weighted edges, with a slight change in treatment. Previous analyses of this data (see, for instance, \cite{gonzalez2025evidence}) showed that negative ties behave quite differently from positive ones. Specifically, there are two main differences. First, negative edges are rarely reciprocated. Typically, one individual reports a negative tie, while the other does not report any relationship at all. We interpret this as an underreporting of negative relations: only one of the people reports the negative tie, but this signals the existence of a conflict regardless of the direction. For this reason, we make no distinction between asymmetric and reciprocal negative ties (reciprocal negative ties are extremely scarce), and simply treat them as indicators of conflict. Second, the distinction between $-1$ and $-2$ ties is weak. Although $-1$ ties are slightly more frequent, both display similar dynamics and degree distributions. We interpret this as evidence that individuals are less precise when judging the intensity of conflicts compared to positive relations. Accordingly, we aggregate both $-1$ and $-2$ ties in our analysis. Finally, note that cases in which one person reports a positive relation and the other a negative one do occur, but they are extremely rare and we do not include them in our analyses. As in the previous section, we compute for each community a set of metrics: raw counts of intra-group and inter-group negative ties; intra-group and inter-group densities of negative ties; the ratio of negative to positive reciprocal ties, for both intra- and inter-group connections; inter-group densities of incoming and outgoing negative ties considered separately; and the average negative in-degree and out-degree. Using these measures, Figure~\ref{fig4} (panels a–e) shows their distributions, separated by group type according to our classification.

Following the previous logic, we ask two main questions: (1) do negative ties appear more often inside or outside groups, and in what proportion? and (2) how frequent are they, inside and outside groups, compared with positive reciprocal ties in the same settings? Panels a), b), and c) address these questions by showing intra- and inter-group densities of positive reciprocal and negative ties, ratios of negative to positive reciprocal ties, and raw counts of internal and external conflicts. As with asymmetric ties, raw counts clearly reveal that external conflicts outnumber internal ones by a large margin. This suggests that conflicts typically occur between groups. At first sight, this aligns with predictions of balance theory at the group level, which argues that conflict should occur between rather than within groups. However, as in the previous section, it is useful to compare these patterns with positive reciprocal ties as a reference, and to examine group types in detail. Ratios of negative to positive reciprocal ties show that, between groups, cohesive communities and sparse peripheries sometimes exhibit more negative than positive reciprocal ties, and sometimes the reverse. In contrast, cores, dense peripheries, and assortative subgroups consistently show lower ratios of negative to positive reciprocal ties externally. In other words, negative ties tend to occur outside groups, but so do positive reciprocal ties, and the latter dominate more strongly when structures are more embedded in larger cohesive structures. This does not necessarily contradict balance theory, as it depends on whether negative ties occur between groups already linked by positive reciprocal ties, or between clusters of groups bound internally by positive ties but opposed to others through negative ties. Still, the picture is far more complex than the clean dichotomies usually assumed in balance theory. In our data, both positive and negative ties co-occur across groups, and we argue that group-level analysis is not the appropriate scale at which to test balance. Instead, local interactions may provide stronger evidence: we argue that in social networks, interactions are often short-ranged and context-driven, and balance is more likely to hold locally, while being violated at medium and long ranges.

Turning to internal conflicts, we observe another interesting pattern. Except in sparse peripheries, reciprocal positive ties greatly outnumber negative ones within groups, as expected. Still, the amount of conflict sustained within groups is notable. The median density of internal positive reciprocal ties is about 0.45, while the median density of internal negative ties is about 0.05, and thus roughly one in ten intra-group ties is a conflict. Breaking this down by group type, cores and assortative subgroups exhibit the lowest ratios, while assortative groups and dense peripheries show higher ratios. As in the case of asymmetric ties, this reveals that groups sustain a substantial number of conflicts. This challenges the idea that groups are defined solely by mutual affinities, and casts doubt on the group-level predictions of balance theory, which assumes that conflict should not occur within groups. It reinforces instead our view that balance should be studied at the local interaction level, not at the group level. One could argue that community detection should explicitly incorporate negative ties, producing smaller subgroups within the groups we identified, in which conflict is externalized and balance may be restored locally. However, we see this as deviating from the plain empirical result. Our groups are identified on the basis of reciprocal positive ties, and thus, even if balance operates locally, cohesive groups of mutual affinities still sustain significant levels of internal conflict, even at the cost of tension.

Finally, we consider the directionality of conflict. Until now we have treated negative ties as undirected, but one might expect some groups to be systematically more `conflictive' (reporting negative ties) and others more `unpopular' (receiving them). Panel d) of Figure~\ref{fig4} shows the distributions of outgoing and incoming negative tie densities by group type, which are very similar. Groups similarly report and receive negative ties. Moreover, conflicts are not evenly distributed across groups: as seen in earlier work, the degree distribution of negative ties is approximately exponential~\cite{gonzalez2025evidence}, with most individuals reporting zero or one conflict and a small number of hubs concentrating the majority of conflicts. Panel e) of Figure~\ref{fig4}, which shows average in- and out-degree for negative ties by group, confirms this. Values are generally similar for incoming and outgoing ties across group types, with no systematic differences. Communities at the upper tail of the distribution likely contain conflict hubs, but these are not tied to any particular group type. Taken together, these findings support our earlier conclusion: groups sustain substantial levels of conflict, and these conflicts tend to cluster around conflictive or unpopular individuals, rather than being evenly distributed across groups.

\section*{Discussion}

In this work, we have characterized the group structure of the personal relationship network of 1068 students from a High School in Madrid. This dataset provides information not only about the structure of friendly relationships, but also about their directionality, their intensity, and the presence of conflictive ties. 
Our inference framework enables us to infer group structure without making strong a priori assumptions about where negative or asymmetric ties should occur and how they should shape group structure.

We first explore a SBM model variant that incorporates all available information about the weights, signs, and directions of ties. We show that while this method works as intended in statistical terms, the resulting clusters do not align with our social intuitions about groupings. The algorithm identifies nodes with similar connectivity patterns, but these patterns are not necessarily meaningful in terms of social behaviour or information. 
This mismatch likely occurs precisely because edge features like direction and sign are included, 
and the resulting complexity tends to obscure interpretability.

To address this issue, we reformulate our strategy to prioritize interpretability of groupings in social terms. We extract the network backbone composed of positive, reciprocal ties, under the assumption that mutual affinities sustain the core of social group structure. On this simplified network, we analyse the resulting community structure. We find that about two-thirds of the communities display assortative patterns, meaning they are more densely connected internally than externally, consistent with typical intuitions about social groupings. However, many of these assortative groups are not isolated but embedded within larger cohesive structures, indicating a dense web of inter-group connections. In addition, core–periphery structures are common, challenging the notion that social group structure is strictly modular. Some nodes also appear relatively isolated and do not fit clearly within the broader community architecture.

Building on this basis, we examine the role and placement of asymmetric and negative ties. We find that asymmetric ties occur predominantly between communities, likely reflecting exploratory or transient interactions beyond the core group. However, a substantial number of asymmetric ties are present within communities, sometimes matching the number of reciprocal ties. This challenges the common assumption that groups are composed solely of mutual affinities. Instead, we argue that group membership may actually stabilize asymmetric ties: while unreciprocated relationships outside groups tend to dissolve, those within groups persist due to social pressure and repeated interaction. Individuals may maintain polite or outwardly positive behaviour that is misinterpreted by others as mutual connection. We also observe that a small fraction of asymmetric ties may reflect hierarchical dynamics, although this is not the dominant pattern. Regarding relationship intensity, both weak and strong reciprocal ties shape internal group structure, but stronger ties play a disproportionately important role relative to their frequency in the overall network.

As for negative ties, we find that they tend to occur between communities, broadly consistent with predictions from balance theory. However, positive reciprocal ties also bridge communities, and these often dominate when groups are embedded in larger cohesive structures. Whether a negative tie represents true inter-group antagonism depends on whether it connects clusters that are otherwise positively linked, or whether it reflects clear group opposition. In our data, both positive and negative ties co-occur across group boundaries. We argue that balance is better assessed at a local scale, as social interactions are often short-ranged and context-driven. Balance may hold in local configurations while being violated at medium or global scales. As with asymmetric ties, we also observe that groups feature a substantial number of negative ties. This contradicts the idea that groups are defined solely by mutual affinity and casts doubt on the typical assumption that conflict does not occur within groups. One might argue that incorporating negative ties into community detection could produce smaller subgroups, externalizing conflict,  but this would lead us away from the empirical basis of our analysis. Our groups are defined through reciprocal positive ties, and our findings show that even within these cohesive structures, conflict is common and persistent. Finally, we find that conflict tends to concentrate around specific individuals (those who are especially unpopular or conflict-prone) rather than being evenly distributed across the network.

Despite having five years of longitudinal data, we did not investigate the dynamics of group evolution. We plan to address this in future work. Some aspects of the dynamic of this dataset have already been explored \cite{escribano2023stability,gonzalez2025evidence}.

All in all, as discussed in the introduction, different social contexts give rise to distinct relational norms that shape group structure. Universal microscopic mechanisms, such as reciprocity, transitivity, balance, or differential popularity, interact with ecological constraints like demographic composition, institutional structure, or educational climate to produce context-specific mesoscopic organization. This microscopic perspective offers a promising avenue for future research. The most compelling questions lie in developing mechanistic explanations that reveal group structure as an emergent property of local interactions, shaped by broader institutional norms and arrangements. In this sense, we advocate for a bottom-up approach (from local interactions to mesoscopic organization) as a necessary complement to the top-down community detection strategy adopted in this study. Moreover, these approaches carry practical relevance. They can inform assessments of school social climate, help identify problematic behavioural patterns, and guide the design of targeted interventions. For instance, the blind application of an SBM that incorporates edge weights, directions, and signs can be useful for detecting clusters of individuals exhibiting similar patterns of conflictual or destabilizing behaviours. Likewise, some of the group types identified in our community characterization, such as sparse peripheries, may serve as proxies for detecting socially vulnerable or isolated individuals. 

As we wrapped up this project we found out about a relevant reference, just appeared on the ArXiv \cite{zhang2025mesoscalestructuressignednetworks}. It also presents an analysis of the mesoscopic structure of signed networks, using different community detection algorithms. In particular, they apply a microcanonical SBM on the full network, as we did in Section A of the SI, finding similar results. Otherwise, the two papers are significantly different.

\section*{Methods}

In this section we provide details about the data collection process, the data composition, the data curation, and the mathematical methods used.
\subsection*{Data collection}
The collection of our data was performed through surveys administered in the school via a computer interface. To extract info about relationships, students were presented with a list of all other students in the high school. They were then asked to select individuals with whom they had a relationship. Specifically, the questionnaire included the following question: `You can now see the list of all the students in the school. Please mark those you have any relationship with by clicking `very good relationship', `good relationship', `bad relationship' or `very bad relationship'. Only one choice is possible. If you do not mark any option, it will be understood that you do not have a relationship with the person'. Typically, it took students about 15 minutes to complete the survey, and they were supervised by a school teacher throughout the process. Our study was approved by the Institutional Review Board of the University Carlos III of Madrid, which stipulated an opt-out procedure. There were no opt-outs, effectively eliminating any potential selection bias. The only students who did not participate were those who were absent on the day of the survey.

\subsection*{Data composition and data curation} 

In Table \ref{table1}, we present the composition of the network, snapshot by snapshot. The Missing Data column includes people who were absent on the day of the survey because there were no opt-outs. We removed some outliers from the analysis, defined as people with more than 30 self-reported very good relationships, more than 50 self-reported good relationships, more than 15 self-reported bad relationships or more than 15 self-reported very bad relationships. These numbers were selected by comparing in-degree and out-degree distributions. The proportion of removed outliers is shown in the Outliers column.

\subsection*{Missing data}
In reconstructing the network snapshots from the survey data, an edge from node \textit{i} to node \textit{j} can be absent for two distinct reasons. First, it may indicate that person \textit{i} deliberately chose not to declare any relationship, positive or negative, with person \textit{j}. Alternatively, the absence of an edge may result from person \textit{i} being absent on the day of the survey, so that the edge is missing simply because no response was recorded. In our analysis, we have carefully distinguished between these two cases, and all missing data are systematically removed from our computations.

\subsection*{Bayesian inference of group structure: Stochastic Block Models}
In this work, we employ a nonparametric Bayesian inference approach to infer group structure from the observed network of personal relationships. A self-contained introduction to these methods can be found in \cite{peixoto2019bayesian}. The framework we use, which includes a family of related models, is generally known as the Stochastic Block Model, or SBM. Here we briefly introduce its basic principles, while referring the reader to the original references for the technical details. All model variants considered in this work were implemented using the Python package graph-tool, version 2.97 \cite{graphtool}.

As mentioned in the introduction, the problem we address is that of inferring an unobserved group structure, given an observed network of connections. Specifically, in our approach we assume the existence of an underlying block structure among the nodes, which plays a role in the generative process of the network, such that the observed network structure is, to some extent, a consequence of this hidden block structure. 
Formally, we assume that nodes are partitioned into $B$ groups, such that each node $i$ belongs to group $b_i$, with $b_i \in [0, B-1]$. We denote the partition by ${\bf b} = \{b_i\}$, i.e., the set of group assignments of all nodes. The specific assumption is that the observed network structure, which we denote by $A$, has been generated given this partition ${\bf b}$. Within this setup, the probability that the observed network $A$ has been generated from any partition ${\bf b}$ is given by
\begin{equation}
P(A \mid \theta, {\bf b}),
\end{equation}

where $\theta$ are additional parameters that control how the partition affects the network structure, allowing us to define different models.

The inference task is therefore to determine the hidden partition ${\bf b}$ from the observed data $A$. By Bayes’ theorem, we can invert the conditional probability above. The models considered here are \textit{microcanonical}, in that the models' constraints encoded with the $\theta$ parameters are imposed strictly, as opposed to only on average, as is typically done. Hence, there exists only one choice of $\theta$ compatible with a given observed network $A$ (given partition ${\bf b}$) and we do not need to sum over $\theta$ throughout computations (more details can be found in \cite{peixoto2017nonparametric}). Thus, by Bayes’ theorem, if we observe a network $A$, the posterior probability that it was generated by a specific partition ${\bf b}$ is
\begin{equation}
P({\bf b} \mid A) = \frac{P(A \mid \theta, {\bf b})  P(\theta, {\bf b})}{P(A)},
\end{equation}
where $P(\theta, {\bf b})$ is the prior probability of the model parameters, and
\begin{equation}
P(A) = \sum_{{\bf b}} P(A \mid \theta, {\bf b}) P(\theta, {\bf b})
\end{equation}
is called the evidence, corresponding to the total probability of the data, summed over all possible parameter values. 

Within this framework, finding the partition that best fits the data corresponds to identifying the partition that maximizes the posterior probability defined above. The framework is quite general, allowing for the specification of many different generative models consistent with a given block structure, by playing with the $\theta$ parameters. This naturally raises the question: How do we select the model that best fits the observed data, while avoiding both overfitting and underfitting? To address this issue, one relies on the concept of description length. In short, the posterior probability can be rewritten as
\begin{equation}
P({\bf b} \mid A) = \frac{\exp(-\Sigma)}{P(A)}.
\end{equation}
With this notation, the description length is defined as
\begin{equation}
\Sigma = - \ln P(A \mid \theta, {\bf b}) - \ln P(\theta, {\bf b}).
\end{equation}
Maximizing the posterior probability (i.e., inferring the most likely partition that explains the observed data given a particular model specification) is equivalent to minimizing this description length. Choosing the model with the smallest description length amounts to selecting the simplest model among all those with comparable explanatory power given the available statistical evidence. 

Again, this framework is highly flexible in defining the specific generative model, leaving room for imposing further constraints. For instance, one may ask: should the nodes follow a specific degree distribution? Should additional generative mechanisms be incorporated beyond the block structure? Should we restrict ourselves exclusively to modular structures? Should we allow for nested modular structures? In order to address these questions, we define a set of generative models that incorporate the different hypotheses and assumptions we hold about the data. In this work we restrict ourselves to the different models included in the graph-tool package. Here, we give conceptual details about the different model variations considered, while referring to the original references for further details. 

\subsubsection*{Variation 1: Standard SBM}
The standard SBM is perhaps the simplest generative model. Its parameters are the partition of the nodes into $B$ groups and a $B \times B$ matrix of edge counts ${\bf e}$, where each entry $e_{rs}$ specifies the number of edges between groups $r$ and $s$. Given these constraints, edges are then placed at random. Consequently, nodes belonging to the same group share the same probability of being connected to other nodes in the network. Notice that this SBM imposes no restrictions on the type of group structure: the edge count matrix ${\bf e}$ is unconstrained. As a result, the model can capture typical assortative communities, where nodes connect primarily within their own group, if this structure best describes the data. However, it can also uncover a wide variety of other structural patterns, such as disassortative (i.e., bipartite), core–periphery, or mixed configurations.

\subsubsection*{Variation 2: Degree-Corrected SBM}
Despite its generality, the standard SBM assumes that edges are placed uniformly at random within groups, consequently nodes in the same group have similar degrees. For networks with heterogeneous degree distributions, a more suitable alternative could be the degree-corrected SBM (DC-SBM) \cite{karrer2011stochastic,peixoto2019bayesian}. This model extends the traditional SBM by including the observed degree sequence of the network as an additional set of parameters. 

\subsubsection*{SBM}
In some cases, networks can exhibit group structure at multiple scales. For example, one might find a clear partition into four large groups, but with additional substructure present within each of them. To capture this, we can employ the nested variation of the SBM, which provides a multilevel hierarchical description of the network \cite{peixoto2014hierarchical}. This model can reveal structural patterns across scales, yielding both coarse and fine-grained representations of the group organization and allowing us to investigate community structure at different levels of granularity.

\subsubsection*{Variation 4: Planted Partition Model}
We may want to impose that the community structure is strictly assortative or disassortative. This extension of the model is known as the planted partition model \cite{zhang2020statistical}.

\subsubsection*{Variation 5: SBM with Triadic Closure}
It is known that a typical generative mechanism in the formation and evolution of social networks is triadic closure. This mechanism favors the formation of an edge between people that share common neighbours. It can lead spontaneously to the formation of a block structure \cite{bianconi2014triadic}. Thus, some block structure may appear, even if it was not there during the generative process of the network, and a simple SBM could find a block structure even if it was not there at the beginning and arose stochastically simply due to the effect of this other generative mechanism. To see if this is the case, we can use a model that incorporates both generative mechanisms \cite{peixoto2022disentangling}, to disregard spontaneous block structure due to the effect of triadic closure from genuine block structure.

\subsubsection*{Model selection}
It is worth mentioning that the models presented here are straightforward variations of the standard SBM. Of course, it is also possible to combine them (for instance, by fitting a nested, degree-corrected SBM). We will take all such combinations into account when selecting the most appropriate model. In all models considered, there is a trade-off. Introducing additional parameters increases the model’s complexity and therefore its description length. However, if the observed data are better explained by this added complexity, the description length may in fact be reduced. To decide which model to use, we compare the description lengths of the fitted models and select the one with the smallest value. Using description length as our comparison metric ensures that the selected model is protected against both overfitting and underfitting. This model selection approach is not the only possible choice within this framework. For instance, instead of selecting the single partition from a model with the smallest description length, we could sample multiple partitions from the posterior distribution and weight them according to their probability, showing that different partitions may explain the observed data with varying likelihoods. However, depending on the level of detail one wishes to develop in the analysis of a single partition, it may not be feasible to examine all sampled partitions at this depth. In our case, we focus on the most probable partition as the explanation of our data, while keeping in mind that it represents the most likely interpretation, not necessarily the only one. 

\subsubsection*{Including directions, weights and signs: layered SBM}
All these models are originally defined for simple networks (undirected, unweighted, and unsigned). However, our network data includes information about the direction, weights, and signs of the edges. Thanks to the flexibility of this framework, it is possible to extend these models to incorporate such information. One approach is to use a layered SBM, in which the edges of the network are distributed across discrete layers, each representing a distinct type of interaction \cite{peixoto2015inferring}. 

\section*{Data availability}
All the aggregated data necessary to replicate our results can be found in this Github repository: \url{TBD}.  

\section*{Competing Interests}
All authors declare no competing interests.

\section*{Acknowledgements}
M.A.G.-C. acknowledges support from the Comunidad de Madrid through the grants for the hiring of pre-doctoral research personnel in training (reference PIPF-2023/COM-29487). M.A.G.-C. and A.S. acknowledge support from grant PID2022-141802NB-I00 (BASIC) funded by MCIN/AEI/10.13039/501100011033 and by ‘ERDFA way of making Europe’, and from grant MapCDPerNets---Programa Fundamentos de la Fundaci\'on BBVA 2022. All authors acknowledge the support of the AccelNet-MultiNet program, a project of the National Science Foundation (Award \#1927425 and \#1927418). This research was supported in part by Lilly Endowment, Inc., through its support for the Indiana University Pervasive Technology Institute.

\section*{Author Contributions}
M.A.G.-C., A.S. and S.F. conceived and conceptualized the research, A.S. collected the data, M.A.G.-C. curated the data, formalized the analyses and obtained the results, and M.A.G.-C., A.S. and S.F. discussed and interpreted the results and wrote the manuscript.

\bibliographystyle{unsrt}
\bibliography{references}  

\newpage 

\begin{table}[t!]
\centering
\caption{Composition of each network snapshot. In the columns Sex, Missing Data and Outliers, numbers represent proportions.}
\label{table1}
\begin{tabular}{lcccc}
Wave & Respondents & Sex (M/F) & Missing Data & Outliers \\
\midrule
DEC 2020 & 409 & 0.52/0.48 & 0.11 & 0.03 \\
MAY 2021 & 409 & 0.52/0.48 & 0.10 & 0.03 \\
SEP 2021 & 530 & 0.49/0.51 & 0.06 & 0.04 \\
FEB 2022 & 530 & 0.49/0.51 & 0.09 & 0.04 \\
MAY 2022 & 530 & 0.49/0.51 & 0.00 & 0.07 \\
SEP 2022 & 524 & 0.54/0.46 & 0.06 & 0.03 \\
JAN 2023 & 535 & 0.54/0.46 & 0.09 & 0.07 \\
MAY 2023 & 536 & 0.54/0.46 & 0.13 & 0.05 \\
SEP 2023 & 554 & 0.51/0.49 & 0.06 & 0.04 \\
JAN 2024 & 565 & 0.51/0.49 & 0.07 & 0.05 \\
MAY 2024 & 566 & 0.51/0.49 & 0.08 & 0.05 \\
SEP 2024 & 553 & 0.52/0.48 & 0.13 & 0.04 \\
JAN 2025 & 567 & 0.52/0.48 & 0.13 & 0.06 \\
MAY 2025 & 564 & 0.53/0.47 & 0.13 & 0.06 \\
\bottomrule
\end{tabular}
\end{table}

\appendix

\section{The blind fit: embedding directions, weights and signs into the SBM}

As discussed in the main text, given that we have detailed information on the directions, weights, and signs of the network links, and that the framework we use is flexible enough to incorporate this information to infer communities, the natural first step is to include all of it within the inference framework. This can be achieved through the layered variant of the SBM, in which the network is represented across four layers, one for each link weight (+2, +1, –1, and –2), and all layers are fitted jointly (see Methods in the main text). The key idea is that the model treats all four layers simultaneously and without imposing any prior assumptions about how each type of link should behave. In doing so, it returns the partition of groups that best explains the structure across all four layers. It is then our task to analyse and interpret which structural features characterize the identified groups. In addition to layering, there are at least two other model features that we should take into account, both in this model and in the other models discussed in the main text.

First, we know that the institutional organization of the high school strongly shapes the network structure. Specifically, students are divided into four courses (i.e., four age groups), within which they attend all classes together, share the same spaces, and spend their free time. As a result, students tend to interact primarily with peers from their own course. This institutional arrangement has an important effect on the community structure: if communities exist, they are likely to display some degree of hierarchical organization. In simple terms, this means that we expect to observe community structure at different scales in the network. At the coarsest level, we anticipate a partition that broadly corresponds to the four courses, while at finer scales we expect additional group structures to emerge within each course. Some community detection algorithms applied to this structure may identify the division into four courses as the best explanation for the data, while overlooking structures at smaller scales. Fortunately, there is a suitable variant of the SBM capable of capturing precisely this type of multiscale organization: the nested SBM (see Methods in the main text). Within the SBM framework, this variant can disentangle the most trivial community division into four courses from the less trivial but equally relevant structures that emerge within them. Another relevant feature to consider in our SBM variant is degree correction. If degree distributions are sufficiently heterogeneous, the model with the smallest description length is expected to be a degree-corrected variant, which incorporates the degree sequence of the nodes as an explicit constraint of the model (see Methods in the main text).

Finally, the layered version of the SBM actually comes in two variants: the independent layers variant and the edge covariates variant. It is not the purpose of this work to go into detail about the differences between these two formulations, but for completeness we included both when considering the set of possible models to explain the observed data. Taking into account all the model features discussed, this results in eight possible models:
\begin{itemize}
    \item Layered SBM (Independent Layers) 
    \item Layered Degree-Corrected SBM (Independent Layers)
    \item Layered Nested SBM (Independent Layers)	 
    \item Layered Nested Degree-Corrected SBM (Independent Layers)
    \item Layered SBM (Edge Covariates) 
    \item Layered Degree-Corrected SBM (Edge Covariates) 
    \item Layered Nested SBM (Edge Covariates) 	 
    \item Layered Nested Degree-Corrected SBM (Edge Covariates) 
\end{itemize}
Interestingly, across all 14 network snapshots, the model that consistently yields the partition with the minimum description length is the Layered Nested Degree-Corrected SBM (Edge Covariates) variant. This model is both nested and degree-corrected. 

In all analyses, both here and in the main text, whenever a nested variant of the SBM is fitted to the data, the algorithm returns a hierarchy of partitions that capture the community structure of the network at different scales. We focus on the most fine-grained division, which we refer to as \textit{partition}, because the divisions at the coarser levels of the hierarchy closely correspond to the division of students by courses, as expected. 

At this point, for each network snapshot, the algorithm returns a partition of the network into communities. It would be tempting to immediately start drawing conclusions about the group structure in networks of personal relationships. However, before doing so, we must ask: what did the algorithm really find? Are these groupings `communities' in the sense we intuitively imagine (e.g., cohesive groups of friends) or are they capturing something else? What aspects of the structure of personal relationships can we infer from this partition? To address these questions,  for each community we measured a set of structural metrics intended to capture its internal organization and external interactions. The aim was to identify what characterizes a community in terms of its observed structural properties. These metrics include:
\begin{itemize}
    \item \textbf{Size:} Number of nodes within the community.
    
    \item \textbf{Density of $+2$ links:} Proportion of $+2$ links within the community.
    
    \item \textbf{Density of $+1$ links:} Proportion of $+1$ links within the community.
    
    \item \textbf{Density of negative links:} Proportion of $-1$ and $-2$ links within the community, treating both as equivalent.
    
    \item \textbf{In- and out-degrees:} Maximum and average in-degree and out-degree of $+2$, $+1$, and negative links for nodes within the community.
    
    \item \textbf{Internal degree fraction:} Maximum and average fraction of each node's in- and out-degree (for $+2$, $+1$, and negative links) that connects to other nodes within the same community.
    
    \item \textbf{Conductance:} Fraction of links incident to community nodes that connect to nodes outside the community (i.e., links with only one endpoint inside the community).
    
    \item \textbf{Density of edge motifs:} Frequency of specific directed dyadic motifs based on tie strength and direction, including:
    \begin{itemize}
        \item \textbf{Intra-group motifs:} \texttt{+2+2}, \texttt{+2+1}, \texttt{+2+0}, \texttt{+1+1}, \texttt{+1+0}, \texttt{+0+0}, \texttt{+0$-1$}.
        \item \textbf{Inter-group motifs:} \texttt{+2+2}, \texttt{+2+1}, \texttt{+2+0}, \texttt{+1+2}, \texttt{+1+1}, \texttt{+1+0}, \texttt{+0+0}, \texttt{+0+1}, \texttt{+0+2}, \texttt{+0$-1$}, \texttt{$-1$+0}.
    \end{itemize}
\end{itemize}

Then, we applied a simple k-means classification algorithm to these communities based on the measured metrics, interpreting each community as a sample, and each metric as a feature of the sample. This allowed us to group together communities with similar profiles. From this, we extracted characteristic patterns of the communities found.

This procedure is neither exhaustive, nor it captures the full richness of the data. Nevertheless, it provides a useful starting point, offering an intuitive first look at what the algorithm has identified as `communities' and helping to guide the deeper analyses that follow.
\begin{figure}[t!]
\centering
\includegraphics[width=\linewidth]{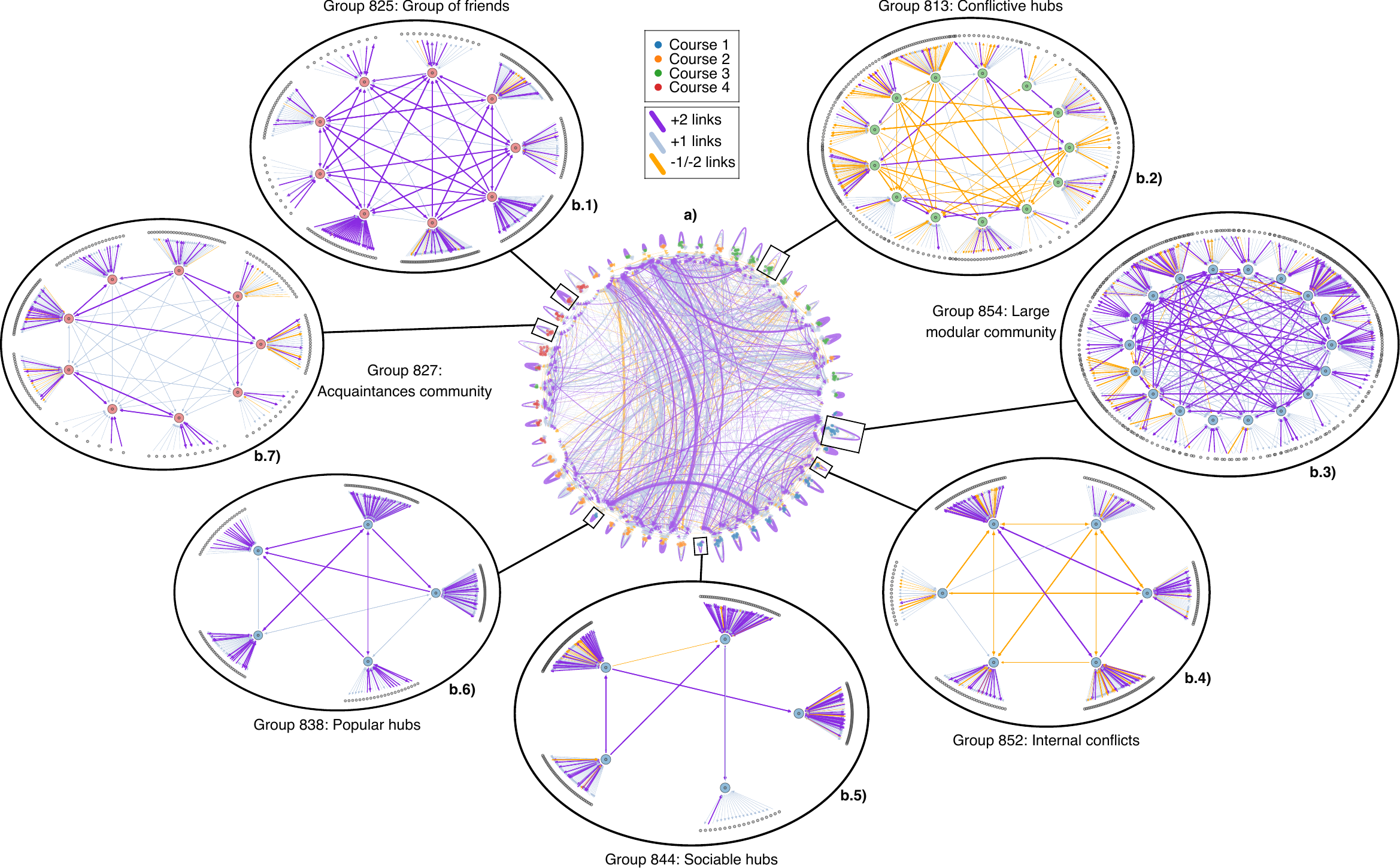}
\caption{\textbf{Panel a)} Network representation of the partition obtained for the May 2023 snapshot, used here as an illustrative example. Nodes, arranged in a circular layout, represent communities, and links represent the connections between them. Each link aggregates all ties of the same weight and direction between individuals belonging to the two corresponding communities, with link width proportional to the total number of aggregated ties. Self-loops represent intra-community connections, while other links correspond to inter-community connections. Link colour indicates the type of tie. Inside each node, we display as many coloured points as individuals assigned to that community, to illustrate differences in community size. The colour of the points indicates the school course of the individuals. Seven communities are highlighted as examples to explore in greater detail in panels b.1)–b.7). \textbf{Panels b.1)–b.7)} Examples of seven communities detected by the algorithm, each illustrating a characteristic structural pattern described in the main text. Here, nodes represent individuals, and links represent ties between them following the colour coding of panel a). Unlike panel a), link widths are uniform since they represent individual rather than aggregated ties. Nodes are arranged in two concentric circles: those in the inner circle belong to the focal community, while those in the outer circle represent individuals outside the community who are connected to its members. This layout allows us to visualize both the internal structure of each community and its external connections to the rest of the network. } 
\label{fig1}
\end{figure}
The algorithm identifies a wide variety of structural patterns, many of which go well beyond the traditional view of assortative communities. To provide a brief illustration, in Figure \ref{fig1} we present the partition obtained for the May 2023 snapshot as an example. The central panel depicts the entangled pattern of connections between communities to illustrate its complexity. From this partition, we highlight seven sample communities that show some of the characteristic structural behaviours uncovered. These communities were randomly selected from their clusters as defined by the k-means classification. The purpose of the figure is not to give an exhaustive description of community structure, but rather to show the diversity of patterns that the algorithm is capturing. As expected, some communities do match an intuitive definition of a group: modular clusters with a high density of positive relationships within them. For example, Community 825 represents a cluster of communities characterized by a large density of reciprocal +2 relationships. However, several other structural behaviours appear that depart from this intuitive picture. Community 852 represents a cluster characterized by a high density of internal conflicts, i.e., nodes grouped together densely connected by negative ties. Nodes are clustered together because they share similar antagonistic behaviours towards each other. Community 813 belongs to a similar cluster, but in this case the communities are defined by the presence of conflictive hubs: nodes with disproportionately large numbers of negative outgoing ties, directed both inside and outside the community. The difference with the cluster represented by Community 852 is that negative ties occur both inside and outside the community, and the characteristic antagonistic behaviour is associated to the nodes themselves even if they do not have a relation to one another. Communities 844 and 838 correspond to groups of sociable or popular hubs. These nodes report or receive many positive ties, respectively, but are not necessarily densely connected among themselves. For example, Community 844 is almost empty internally, most of its ties point outwards. By contrast, Communities 827 and 854 display more familiar “cohesive” group structures, similar to Community 825. The difference between the three of them is the characteristic behaviour of some internal characteristics, like the density of +1 ties or the group size, but they still resemble intuitive social clusters.

From this example it becomes clear that, if our goal is to analyse group structure in the social sense described in the introduction (groups shaped by personal affinities and shared membership as we intuitively conceive them) then we cannot simply fit an SBM blindly using all available information. The algorithm itself is working exactly as intended: it summarizes the network by grouping nodes whose connectivity patterns are statistically alike. The issue is not with the method, but with the mismatch between the structural groupings inferred by the algorithm and the social groupings we aim to study and the interpretation we give them. This result motivates a reframing of our strategy, which we explore in the main text.

\newpage 

\section{Approximation for the group type characterization}
As explained in the main text, the SBM imposes no restrictions on the type of block structure inferred, so the edge count matrix is unconstrained. It can therefore encode a variety of patterns (assortative, disassortative, core–periphery, and combinations thereof). Given this matrix, we want to determine the nature of each group and understand why the algorithm identified it as a community. To fully characterize a group $r$, one would need to examine $e_{rr}$ (links within $r$) together with all $e_{rs}$ (links from $r$ to every other group). For example, if $e_{rr}$ is large while all $e_{rs}$ are small, $r$ is a typical assortative community. But if $e_{rs}$ is large for some $s$, $r$ may in fact constitute the periphery of group $s$. The scenario is more complex when $r$ shows both a strong internal edge density and significant connections to multiple other groups also strongly connected among themselves.

To simplify the problem, we adopt an approximation. For each group $r$, we focus only on three entries of the edge count matrix: $e_{rr}$, $e_{r\bar{s}}$, and $e_{\bar{s}\bar{s}}$, where $\bar{s}$ is the group to which $r$ is most strongly connected (i.e., the $s\neq r$ that maximizes $e_{rs}$). The rationale is that, as a first approximation, the structural role of $r$ can be characterized by looking at its internal density and the group it is most strongly connected to. In general, if communities tend to be highly interconnected and community $r$ maintains strong ties with several others, this would be a poor approximation, since focusing only on a subset of its connections would ignore much of the relevant structure. However, if community $r$ concentrates most of its links internally and with just one or a few other communities, then characterizing it using only $e_{rr}$, $e_{r\bar{s}}$, and $e_{\bar{s}\bar{s}}$ is a reasonable approach. 

\begin{figure}[h!]
\centering
\includegraphics[width=0.5\linewidth]{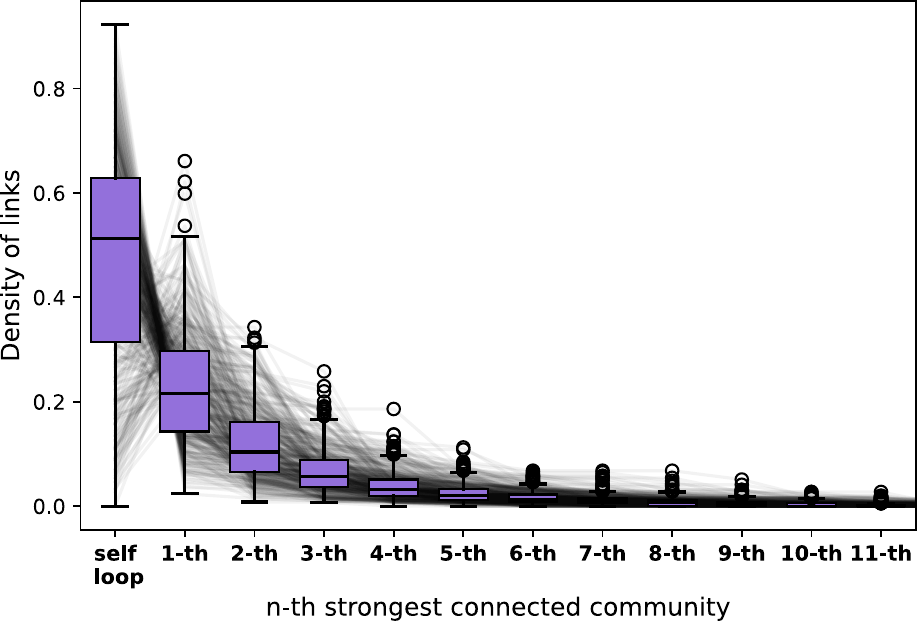}
\caption{Distributions of the density of intra-community links of the focal community and inter-community links between the focal community and the others, ranked from the community most strongly connected to the focal one to the community most weakly connected to it. The first position on the x-axis represents the density of intra-community links (self-loop of the focal community), followed by the inter-community densities ranked from the strongest to the weakest. Black lines represent individual behaviours, while the purple box plots summarize their distributions.} 
\label{entropy}
\end{figure}

To determine if this approximation is appropriate in our case, for each focal community we compute the intra-group edge density and the densities of edges between the focal community and each other community, ranked from the strongest connected to the weakest connected community. In Figure \ref{entropy} we illustrate the distributions of these quantities. The average pattern follows a fast decaying behaviour: most links are intra-community, while inter-community links are largely concentrated toward a single other community rather than evenly spread across many. In some cases, intra-community links are less dominant, but links are then still concentrated towards the most strongly connected community. The focal community and the one it is most strongly connected to capture roughly 75-80\% of all connections, while the remaining links are distributed in very small amounts across the other groups. Focusing only on external connections, almost 50\% of them belong to the most strongly connected community, while the remaining links are again distributed in small amounts across the rest. This result provides justification for our approximation.

\end{document}

\end{document}